\documentstyle[preprint,pra,aps]{revtex}
\newcommand{\asym}[1]{\hspace{7pt}\raisebox{.0ex}{$\simeq $}\hspace{-12pt}
\raisebox{-1.3ex}{$_{#1}$}\hspace{3pt}}

\begin{document}
\draft
\title{Multiphoton detachment of electrons from negative ions}
\author{G. F. Gribakin\cite{ggem} and M. Yu. Kuchiev\cite{mkem}}
\address{School of Physics, University of New South Wales,
Sydney 2052, Australia}
\maketitle
\begin{abstract}
A simple analytical solution for 
the problem of multiphoton detachment from negative ions
by a linearly polarized laser field is found.  It 
is valid in the wide range of intensities and frequencies of the field,
from the perturbation theory to the tunneling regime, and is applicable to
the excess-photon as well as near-threshold detachment.
Practically, the formulae are valid when the number of photons is
greater than two. They produce the total detachment rates,
relative intensities of the excess-photon peaks, and photoelectron angular
distributions for the hydrogen and halogen negative ions, in agreement
with those obtained in other, more numerically involved calculations in
both perturbative and non-perturbative regimes. Our approach explains the
extreme
sensitivity of the multiphoton detachment probability to the asymptotic
behaviour of the bound-state wave function. Rapid oscillations in
the angular dependence of the $n$-photon detachment probability are
shown to arise due to interference of the two classical trajectories
which lead to the same final state after the electron emerges at the
opposite sides of the atom when the field is close to maximal.
\end{abstract}
\newpage
\section{Introduction}\label{intr}
In this paper we present an analytical solution to the problem of
multiphoton detachment from a negative ion by a linearly polarized laser
field. It gives very reliable quantitative results for a wide range of
intensities and frequencies of the laser field, from the weak-field regime,
where the process is described by the perturbation theory, to the
strong fields where it proceeds as tunneling. The theory
is valid when the number of photons $n$ is large, but usually gives good
results as soon as $n>2$. We use it to calculate and examine various
characteristics of the problem: the total multiphoton detachment rate, the
$n$-photon detachment cross sections, the spectrum of excess-photon
detachment (EPD) photoelectrons (the analogue of above-threshold
ionization in atoms), and the peculiar photoelectron angular distributions.

There are two important physical properties of the multiphoton detachment
process:\\
(i) The frequency of the laser field is much lower than the electron
binding energy,
\begin{equation}\label{omllE}
\omega \ll |E_0|~,
\end{equation}
where $E_0=-\kappa ^2/2$ is the energy of the bound state (atomic units
are used throughout). This means that multiphoton detachment is an
{\em adiabatic problem}. The external field varies slowly in comparison
with the period of electron motion in the system. Therefore, the general
adiabatic theory \cite{dykhne,landau,keldysh} is applicable. As long as
the laser field is weaker than the atomic field, the detachment
probability is exponentially small with respect to the adiabaticity
parameter $|E_0|/\omega \sim n$.\\
(ii) The process of multiphoton detachment takes place when the
electron is far away  from the atomic particle (see section
\ref{th}), at {\em large distances}
\begin{equation}\label{rlarge}
r\sim R=\sqrt{\frac{\gamma }{\omega \sqrt{1+\gamma ^2}}} \gg 1~,
\end{equation}
where $\gamma =\omega \kappa /F$ is the Keldysh parameter and $F$ is the
field strength. In the weak field regime, $\gamma \gg 1$, Eq.
(\ref{rlarge}) gives
$R\simeq 1/\sqrt{\omega}\sim \kappa ^{-1}\sqrt{2n}\gg 1$, where
$\kappa \sim 0.3$ for a typical negative ion binding $|E_0|\sim 1$~eV. In
the strong field regime, $\gamma <1$, estimate (\ref{rlarge}) yields
$R\simeq \sqrt{\gamma /\omega }=\kappa ^{-1}\sqrt{F_0/F}
\gg 1$, where $F_0\equiv \kappa ^3$ is the typical atomic electric field,
$F\ll F_0$.

The two features (i) and (ii) greatly simplify the multiphoton
detachment problem. Owing to (ii), the final state of the electron
can be described by the Volkov wave function \cite{Wolk}
which takes into account the external field and neglects the atomic field.
Moreover, the Volkov wave function describes explicitly the variation of
the electron energy in the laser field. This makes it very convenient for
application of the general adiabatic theory, as suggested by (i).

Calculations based on the Volkov final-state wave function were first
done by Keldysh \cite{keldysh}. Subsequently, the idea was developed by
Perelomov {\em et el} \cite{perelom} and later reconsidered by Faisal
\cite{faisal} and Reiss \cite{reiss}.
This approach is usually supposed to give a correct qualitative picture of
multiphoton processes. In this paper we demonstrate that, in fact, it
produces very accurate quantitative results for the multiphoton
detachment from negative ions. We re-examine and extend the
Keldysh theory, paying particular attention to the following points. 
First, we show that the EPD can be accurately described by the theory.
Originally, the theory was developed for low-energy photoelectrons
\cite{keldysh} with kinetic energies much smaller than
the binding energy. The present approach is valid at any photoelectron
energy. Secondly, the angular distribution of photoelectrons is
examined in detail. We show that a nontrivial oscillatory pattern of the
angular distribution is caused by the simple and interesting physics.
The photoelectron's escape from the atomic particle is most probable
when the field reaches its maximum. There are two such instants in every
period of the laser field $T= 2\pi/\omega$, say, $t=0$ and
$t=T/2$. As a result, there are two classical trajectories which
lead to the same final state of the photoelectron. Interference of the
corresponding amplitudes gives rise to an oscillatory angular dependence
of the detachment rate. There is a similar effect in the single-photon
detachment in the presence of a static electric field, where the
interference takes place between the two trajectories of the electron
emitted up or down field \cite{fabrikant,demkonos}.

Estimate (\ref{rlarge}) leads to a further simplification of the problem,
since the initial bound-state wave function of the atomic system should
also be considered at large distances, where it can be replaced by its
simple asymptotic form. The complicated behaviour of the wave function
inside the atom, and the corresponding many-electron dynamics have little
influence on the multiphoton detachment. In contrast, use of the wave
function with incorrect asymptotic behaviour, e.g. that corresponding to
the Hartree-Fock binding energy, introduces an error, which is
exponentially large with respect to $\sqrt{n}$. Such sensitivity
has been noticed in the perturbation theory calculations of the two- and
three-photon detachment from H$^-$ \cite{star}.

There has been a large number of papers where multiphoton detachment
from the hydrogen and halogen negative ions is investigated. Perturbation
theory calculations include those based on the Hartree-Fock approximation
\cite{crance}, adiabatic hyperspherical approach \cite{star}, model
potential \cite{chu}, a configuration-interaction procedure \cite{hart94},
and the Lippmann-Schwinger equation \cite{proulx}.
There are also numerous non-perturbative methods, such as the Floquet
close-coupling method \cite{dimou}, complex-scaling generalized
pseudospectral method \cite{wang}, non-Hermitian Floquet Hamiltonian method
\cite{telnov}, and the $R$-matrix Floquet theory \cite{dorr,hart96}.
All the above methods rely on much more involved numerical calculations
than those needed in our analytical approach. However, we believe that
the present theory provides accurate answers for most of the multiphoton
detachment problems. For illustration purposes, we reproduce a variety of
results obtained earlier, including the $n$-photon cross sections, total
detachment probability, EPD spectrum and photoelectron angular
distributions for a large range of frequencies and intensities of the
field (Sec. \ref{ex}). We believe that in some cases our results are more
accurate than those obtained previously, due to the correct asymptotic
behaviour of the bound-state wave function we use.

The good accuracy we have achieved within the Keldysh-type theory is
quite useful for the multiphoton detachment problem. On the other hand,
its validity is very important for the development of an adiabatic
theory of more complicated phenomena, such as double ionization
\cite{kuchiev87,kuchiev95,kuchiev96}.

The formulae obtained in this paper can be used to estimate probabilities
of multiphoton ionization of neutral atoms. However, the influence of
the Coulomb field of the positive ion on the wave function of the
photoelectron cannot be neglected \cite{perpop,zaret}, and our results for
the multiphoton ionization would be less reliable.

\section{Theory}\label{th}

\subsection{Basic equations}\label{baseq}

Consider the removal of a valence electron from an atom or a negative ion
by the laser field ${\bf F} (t)={\bf F}\cos \omega t$. The differential
detachment rate can be written as the sum over $n$-photon processes [see
Appendix \ref{prob}, Eq. (\ref{fincont})],
\begin{equation}\label{difprob}
dw_n=2\pi \sum _n|A_{{\bf p}n}|^2 \delta(E_{\bf p}-E_0-n\omega )
\frac{d^3p}{(2\pi )^3}~,
\end{equation}
where $A_{{\bf p}n}$ is the amplitude of the $n$-photon process,
\begin{equation}\label{Apn}
A_{{\bf p}n}=\frac{1}{T}\int _0^T\Psi ^*_{\bf p}({\bf r},t)V_F(t)
\Psi _0({\bf r},t)d{\bf r}dt~,
\end{equation}
$\Psi _0({\bf r},t)=e^{-iE_0t}\Phi _0({\bf r})$ is the wave function
of the initial electron state in the atomic potential $U({\bf r})$,
\begin{equation}\label{Phi0}
\left[ \frac{p^2}{2}+U({\bf r})\right] \Phi _0({\bf r})=
E_0\Phi _0({\bf r})~,
\end{equation}
$V_F(t)$ is the interaction with the laser field,
\begin{equation}\label{VF}
V_F(t)=-e{\bf r}\cdot {\bf F}(t)~,
\end{equation}
in the length gauge, $e=-1$ for the electron,
and $\Psi _{\bf p}({\bf r},t)$ is the continuous spectrum solution of
the time-dependent Shr\"odinger equation with the quasienergy
$E_{\bf p}={\bf p}^2/2+F^2e^2/4\omega ^2$. It describes the outgoing
photoelectron in the laser field with the translational momentum ${\bf p}$,
and $F^2e^2/4\omega ^2$ is the electron quiver energy due to the field.
The subscript $n$ in $A_{{\bf p}n}$ reminds one that the amplitude must be
calculated at $E_{\bf p}=E_0+n\omega $ provided by the energy conservation
in Eq. (\ref{difprob}).

As we show below, the detachment probability
is determined by the asymptotic behaviour of the bound-state wave function
at large distances. This means that the role of electron correlations
in the multiphoton detachment of a single electron is small, provided
$\Phi _0({\bf r})$ represents correctly the asymptotic behaviour of the
true many-electron wave function of the system,
\begin{equation}\label{mewf}
\Psi _N({\bf r}_1,\dots ,{\bf r}_{N-1},{\bf r})\asym{r\gg 1}
\Psi _{N-1}({\bf r}_1,\dots ,{\bf r}_{N-1})\Phi _0({\bf r})~,
\end{equation}
where $\Psi _N$ is the ground-state wave function of the $N$-electron
system, and $\Psi _{N-1}$ is the wave function of the $N-1$-electron atomic
residue.

If we neglect the influence of the atomic potential $U({\bf r})$ on the
photoelectron, the final-state is given by the Volkov wave function,
\begin{equation}\label{volkov}
\Psi _{\bf p}({\bf r},t)=\exp \left[ i({\bf p}+{\bf k}_t)\cdot {\bf r}
-\frac{i}{2}\int ^t({\bf p}+{\bf k}_{t'})^2dt'\right] ~,
\end{equation}
where ${\bf k}_t=e\int ^t{\bf F}(t')dt'$ is the
classical electron momentum due to the field. By omitting the lower
integration limit we mean that we set its contribution to zero, as if
the integration is performed from $-\infty $ and the integrand is switched
on adiabatically. For the Volkov function (\ref{volkov}) this gives the
same phase as in \cite{kuchiev95}, $-\frac{i}{2}\int _0^t
({\bf p}+{\bf k}_{t'})^2dt'+i{\bf p}{\bf F}/\omega ^2$, and provides
$\Psi _{\bf p}({\bf r},t)$ with a convenient symmetry property with respect
to inversion:
\begin{equation}\label{symvolk}
\Psi _{-{\bf p}}({\bf r},t)=\Psi _{\bf p}(-{\bf r},t+T/2)
\exp (iE_{\bf p}T/2)~.
\end{equation}
The wave function (\ref{volkov}) satisfies the Shr\"odinger equation
\begin{equation}\label{tdS}
i\frac{\partial \Psi _{\bf p}}{\partial t}=\left[ \frac{p^2}{2}
+V_F(t)\right] \Psi _{\bf p}~.
\end{equation}
The neglect of the short-range potential $U({\bf r})$ for
the photoelectron is justified in multiphoton processes, e.g., in the
multiphoton detachment from negative ions, see end of Sec. \ref{adap}.

Using the complex conjugates of Eqs. (\ref{volkov}) and (\ref{tdS}), and
$i\partial \Psi _0/\partial t=E_0 \Psi _0$, we transform amplitude
(\ref{Apn}) into
\begin{equation}\label{Apn1}
A_{{\bf p}n}=\frac{1}{T}\int _0^T \left[E_0-\frac{({\bf p}+{\bf k}_t)^2}
{2}\right] \tilde \Phi _0({\bf p}+{\bf k}_t)
\exp \left[\frac{i}{2}\int ^t({\bf p}+{\bf k}_{t'})^2dt'-iE_0t \right]
dt~,
\end{equation}
where $\tilde \Phi _0({\bf q})$ is the Fourier transform of
$\Phi _0({\bf r})$,
\begin{equation}\label{Fou}
\tilde \Phi _0({\bf q})= \int d{\bf r}e^{-i{\bf q}\cdot {\bf r}}
\Phi _0({\bf r})~.
\end{equation}

Note that in the velocity gauge,
\begin{equation}\label{Vgauge}
V_F(t)=-\frac{e}{c}{\bf A}(t)\cdot
{\bf p}+ \frac{e^2}{2c^2}{\bf A}^2(t)~, \qquad {\bf A}(t)=-c\int ^t
{\bf F}(t')dt'~,
\end{equation}
the Volkov wave function looks simpler,
\begin{equation}\label{volkovV}
\Psi _{\bf p}({\bf r},t)=\exp \left[ i{\bf p}\cdot {\bf r}
-\frac{i}{2}\int ^t({\bf p}+{\bf k}_{t'})^2dt'\right] ~.
\end{equation}
This gauge, which apparently `leads to an analytical simplicity'
\cite{reiss},
\begin{equation}\label{Apn2}
A_{{\bf p}n}=\frac{1}{T}\left( E_0-\frac{{\bf p}^2}{2}\right)
\tilde \Phi _0({\bf p}) \int _0^T 
\exp \left[\frac{i}{2}\int ^t({\bf p}+{\bf k}_{t'})^2dt'-iE_0t \right]
dt~,
\end{equation}
is though less physical than the length gauge in this problem.
The amplitude (\ref{Apn}) is not gauge invariant when $U({\bf r})$
is neglected for the final state (compare (\ref{Apn1})
with (\ref{Apn2})), except for the zero-range $s$-wave initial state,
$\Phi _0({\bf r})=A e^{-\kappa r}/\sqrt{4\pi }$. The length gauge
interaction (\ref{VF}) emphasizes large distances, where the
bound-state wave function $\Phi _0({\bf r})$ has a well-defined asymptotic
behaviour. We will see in the next section that this gives it a major
advantage over the velocity gauge. In the limit $\omega \rightarrow 0$
the length-gauge calculation reproduces the static-field result
\cite{demkov,smch}.

\subsection{Adiabatic approximation}\label{adap}

For multiphoton processes the integral over time in the amplitude
(\ref{Apn1}), contains a rapidly oscillating exponent
$\exp [iS(\omega t)]$, where $S(\omega t)\sim 2\pi n$ is the
coordinate-independent part of the classical action
\begin{equation}\label{action}
S(\omega t)=\frac{1}{2}\int ^t({\bf p}+{\bf k}_{t'})^2dt'-E_0t ~.
\end{equation}
This makes the amplitude $A_{{\bf p}n}$ exponentially small, and the
integral $\int _0^T\dots dt$ should be calculated using the saddle-point
method. The positions of the saddle points are given by $dS(wt)/dt=0$,
which yields
\begin{equation}\label{speq}
({\bf p}+{\bf k}_{t})^2+\kappa ^2=0~.
\end{equation}
The saddle-point method in this problem has a simple and important
physical contents.
The two terms in the right-hand side of Eq. (\ref{action}) describe the
energy of the electron in the initial and final states, $E_0$ and
$({\bf p}+{\bf k}_t)^2/2$, respectively. According to the
general adiabatic theory \cite{landau}, the transition from the initial
into the final state happens at the moment of time when their energies
are equal. This is exactly the meaning of Eq. (\ref{speq}).

Note that condition (\ref{speq}) coincides with the positions of
singularities of the Fourier transform $\tilde \Phi _0({\bf p}+{\bf k}_t)$
in the amplitude (\ref{Apn1}). Indeed,
the general asymptotic form of $\Phi _0({\bf r})$ is
\begin{equation}\label{asymp}
\Phi _0({\bf r})\asym{r\gg 1} Ar^{\nu -1}\exp (-\kappa r)
Y_{lm}(\hat {\bf r})~,
\end{equation}
where $\nu =Z/\kappa $, $Z$ is the charge of the atomic residue
($\nu =Z=0$ for the negative ion), and $\hat {\bf r}={\bf r}/r$ is the
unit vector. It is easy to see
that due to (\ref{asymp}) the Fourier transform (\ref{Fou}) is singular at
$q^2=-\kappa ^2$. Using \cite{ryzhik} we derive the following asymptotic
form of $\tilde \Phi _0({\bf q})$ for $q\rightarrow \pm i\kappa $,
\begin{equation}\label{asymF}
\tilde \Phi _0({\bf q})\simeq 4\pi A(\pm )^lY_{lm}(\hat {\bf p})
\frac{(2\kappa )^\nu \Gamma (\nu +1)}{(q^2+\kappa ^2)^{\nu +1}}~,
\end{equation}
where $(\pm )^l\equiv (\pm 1)^l$ corresponds to
$q\rightarrow \pm i\kappa $.

Therefore, when the
length-form amplitude is calculated by the saddle-point method, we do not
need to know the behaviour of the bound-state wave function in the whole 
space. In contrast, when using the velocity-form amplitude (\ref{Apn2}),
the value of the Fourier transform for the true final-state momentum
${\bf p}$ is needed. To calculate it one must know the exact wave function
at all distances including $r\sim 1$. What makes the problem even harder,
many-electron correlations become essential there.

Equation (\ref{speq}) for the saddle-points presented explicitly as
\begin{equation}\label{speq1}
\left( {\bf p}+\frac{e{\bf F}}{\omega }\sin \omega t_\mu \right) ^2
+\kappa ^2=0
\end{equation}
defines complex values $t_\mu $ where the transition from the bound
state into the Volkov state takes place. Equation (\ref{speq1}) has two
pairs of complex conjugate roots in the interval
$0\leq {\rm Re}\, (\omega t )<2\pi $. According to the general theory of
adiabatic transitions \cite{landau}, in the case when the final-state
energy $E_{\bf p}$ is greater then the initial energy $E_0$, we should
take into account the two saddle-points in the upper half-plane of complex
$t$.

Changing the integration variable to $\omega t$ and substituting the
asymptotic expression for the Fourier transform near the singularity
(\ref{asymF}), we can write amplitude (\ref{Apn1}) as the sum over the two
saddle points,
\begin{equation}\label{Asad1}
-\frac{1}{2\pi }\sum _{\mu=1,2}
\int [({\bf p}+{\bf k}_t)^2+\kappa ^2]\frac{4\pi A(\pm )^l
Y_{lm}(\hat {\bf p}_\mu ) (2\kappa )^\nu \Gamma (\nu +1)}
{2[({\bf p}+{\bf k}_t)^2+\kappa ^2]^{\nu +1}}\exp [iS(\omega t)]
\,d(\omega t)~,
\end{equation}
where the integral is taken over the vicinity of the $\mu $th saddle point,
$\hat {\bf p}_\mu $ is the unit vector in the direction of
${\bf p}+(e{\bf F}/\omega )\sin \omega t_\mu $, and the two signs in
$(\pm )$ correspond to $\mu =1,\,2$. Note that for the initial electron
state bound by short-range forces, as in a negative ion,
the integrand in (\ref{Asad1}) has no singularity ($\nu =0$), and the
application of standard saddle-point formulae is straightforward.
Having the general case in mind, we will calculate the amplitude for
arbitrary $\nu $, taking into account the singularity at the saddle point. 
This is also useful if one wants to calculate the amplitude in the original
form (\ref{Apn}), without using the transformation which leads to Eq.
(\ref{Apn1}).

Using $dS(\omega t)/d(\omega t)=[({\bf p}+{\bf k}_t)^2+\kappa ^2]/2\omega$,
we can re-write Eq. (\ref{Asad1}) as
\begin{equation}\label{Asad2}
-2\pi A\Gamma (\nu +1)\left( \frac{\kappa}{\omega }\right) ^\nu
\frac{1}{2\pi }\sum _{\mu=1,2}(\pm )^l Y_{lm}(\hat {\bf p}_\mu )
\int \frac{\exp [iS(\phi )]}{[S'(\phi )]^\nu }d\phi ~,
\end{equation}
where $\phi =\omega t$. In the vicinity of the saddle point $\phi _\mu $,
$S'(\phi _\mu )=0$, we have
$S'(\phi )\simeq S''(\phi _\mu )(\phi -\phi _\mu )$.
The contribution of this saddle point is then given by the following
integral
\begin{equation}\label{vicin}
\int \frac {\exp [iS(\phi )]}{[S'(\phi )]^\nu }d(\phi )=
\frac {1}{[S''(\phi _\mu )]^\nu }\int \frac {\exp [iS(\phi )]}
{(\phi -\phi _\mu )^\nu }d\phi ~,
\end{equation}
which is calculated in Appendix \ref{apsad}.

The explicit form of the action (\ref{action}) is
\begin{equation}\label{acexp}
S(\phi )=n\phi -\xi cos \phi -\frac{z}{2}\sin 2\phi ~,
\end{equation}
where $z=e^2F^2/4\omega ^3$ is the mean quiver energy of the electron in
the laser field in units of $\omega $, $\xi =e{\bf Fp}/\omega ^2$ depends
on the angle $\theta $ between the photoelectron momentum ${\bf p}$ and the
field ${\bf F}$, and we put $E_{\bf p}-E_0=n\omega$ due to the
energy conservation in (\ref{difprob}). Thus, we obtain the
final expression for the amplitude by the saddle-point method,
\begin{eqnarray}\label{Asad3}
A_{{\bf p}n}=-2\pi A \Gamma (1+\nu /2) 2^{\nu /2}
\left( \frac{\kappa}{\omega }\right) ^\nu
\sum _{\mu=1,2}(\pm )^l Y_{lm}(\hat {\bf p}_\mu )
\frac{(c_\mu +is_\mu )^n\exp [ -ic_\mu (\xi +zs_\mu )]}
{\sqrt{2\pi (-iS''_\mu )^{\nu +1}}}~,
\end{eqnarray}
where
\begin{eqnarray}\label{smu}
&&\sin \omega t_\mu =(-\xi \pm i\sqrt{8z(n-z)-\xi ^2})/4z\equiv s_\mu  ,\\
&&\cos \omega t_\mu =\pm \sqrt {1-s_\mu ^2}\equiv c_\mu ~,\label{cmu}\\
&&S''_\mu = c_\mu (\xi +4zs_\mu ) ~,\label{Smu}
\end{eqnarray}
and the signs $\pm $ correspond to the two saddle points $\mu =1,2$.
The usual definition of the spherical harmonics \cite{varsh}
\begin{equation}\label{Ylm}
Y_{lm}(\vartheta ,\varphi )=
\frac{1}{\sqrt{2\pi}}e^{im\varphi }(-1)^{\frac{m+|m|}{2}}
\left[ \frac{2l+1}{2}\,\frac{(l-|m|)!}{(l+|m|)!}\right] ^{1/2}P_l^{|m|}
(\cos \vartheta )~,
\end{equation}
is generalized naturally to calculate $Y_{lm}(\hat {\bf p}_\mu )$ for
complex vectors by setting
\begin{equation}\label{cos}
\cos \vartheta =\frac{({\bf p}+{\bf k}_t)\cdot {\bf F}}
{\sqrt{({\bf p}+{\bf k}_t)^2}F}=\sqrt{1+\frac{p_\perp ^2}{\kappa ^2}}~
\end{equation}
where the last equality is valid at the saddle points, $p_\perp $ is the
component of ${\bf p}$ perpendicular to ${\bf F}$, $p_\perp =p\sin
\theta $. The real physical angle $\theta $ should not be confused with
the complex angle $\vartheta $ from equations (\ref{Ylm}) and (\ref{cos}).
The azimuthal angle $\varphi $ is the same in both cases.

Using (\ref{symvolk}) and the symmetry of the spherical harmonics $Y_{lm}$
one can show that the amplitude (\ref{Apn}) has the following symmetry
properties: $A_{{\bf p}n} \rightarrow (-1)^{n+l}A_{{\bf p}n}$ upon
inversion ${\bf p}\rightarrow -{\bf p}$ ($\theta \rightarrow \pi -\theta $,
$\varphi \rightarrow \varphi +\pi $), and
$A_{{\bf p}n}\rightarrow (-1)^{n+l+m}A_{{\bf p}n}$, upon reflection in the
plane perpendicular to the direction of the field
($\theta \rightarrow \pi -\theta $). Consequently, the amplitude is zero
for ${\bf p}$ perpendicular to the field, if $n+l+m$ is odd.

It is easier to look at the physics behind Eqs. (\ref{Asad3})-(\ref{Smu})
in the case when the photoelectron momentum is small, $p \ll \kappa $. The
following simpler formulae for the saddle points can be obtained from
(\ref{smu})--(\ref{Smu}) by setting $\xi =0$,
\begin{equation}\label{gamma}
\sin \omega t_\mu =\pm i\gamma ,\quad
\cos \omega t_\mu =\pm \sqrt {1+\gamma ^2} ,\quad
S''_\mu = i\gamma \sqrt{1+\gamma ^2}\frac{F^2}{\omega ^3}~,
\end{equation}
where $\gamma =\kappa \omega /F$ is the Keldysh parameter. Thus, for small
photoelectron momenta the saddle points are
$\omega t_1=i\sinh ^{-1}\gamma $ and $\omega t_2=\pi+i\sinh ^{-1}\gamma $,
and the detachment takes place at the two
instances when the external field is maximal, $t=0$ and $T/2$ on the real
axis. Accordingly, the total amplitude (\ref{Asad3}) is the sum of the two
contributions from these points. This results in oscillations in the
photoelectron angular distribution, which we discuss in greater detail
below.

The original approach used in \cite{keldysh,perelom} was to
expand Eqs. (\ref{smu}), (\ref{cmu}) and the action (\ref{acexp}) in powers
of $p/\kappa $ to the second order (see Sec. \ref{lowen}), thus obtaining
corrections to (\ref{gamma}). In this regime $\gamma $ remains the main
parameter which determines the probability of multiphoton detachment
\cite{note}. However, the applicability of the saddle-point result
(\ref{Asad3}) is essentially narrowed by such expansion (Sec. \ref{ex}).

The adiabatic nature of the problem allows us to estimate the radial
distances which are important in the multiphoton detachment process. We
have already seen
that the saddle points in the integral in (\ref{Apn1}) coincide with the
poles of the Fourier transform $\tilde \Phi _0({\bf q})$. The form of
$\tilde \Phi _0({\bf q})$ at $q\rightarrow \pm i\kappa $ is given by
the behaviour of $\Phi _0({\bf r})$ at $r\rightarrow \infty $. To estimate
the essential distances
look at Eq. (\ref{Asad2}). The range of $\phi $ where the integral is
saturated is determined by $|S''(\phi _\mu )(\delta \phi )^2|\sim 1$,
which gives $\delta \phi \sim |S''(\phi _\mu )|^{-1/2}$. The
corresponding range of momenta ${\bf p}+{\bf k}_t$ is given by
\begin{equation}\label{dq}
\delta q\sim \frac{F}{\omega }\cos \phi _\mu \,\delta \phi \sim
\left( \frac{\omega \sqrt{1+\gamma ^2}}{\gamma }\right) ^{1/2}
\equiv \frac{1}{R}~,
\end{equation}
where we use Eqs. (\ref{gamma}). The essential distances are obtained from
$r\,\delta q\sim 1$, which yields estimate (\ref{rlarge}). It is important
that $R\gg 1$ in both weak- and strong-field
regimes. This makes the Keldysh approach valid for
short-range potentials. There is another physical reason which helps to
understand why the atomic potential can be neglected for the photoelectron.
When a large number of photons is absorbed by the photoelectron, higher
angular momentum partial waves are populated. The influence of the
short-range potential upon them is small.
For a given electron momentum $p$ the important $l$ values can be estimated
as $l\sim pR$. In the perturbation theory regime this estimate yields
$l\sim (p/\kappa )\sqrt{n}$, which suggests that the spread of the
probability to find the photoelectron with given $l$ is described by a
random walk of $n$ steps.

Estimate (\ref{rlarge}) also explains the extreme sensitivity of the
multiphoton detachment rates to the asymptotic behaviour of the
bound-state wave function. Suppose a bound state wave function
characterised by $\kappa '$ instead of the true $\kappa $ is used. The
error in the amplitude (\ref{Apn}) introduced by replacing $\kappa $ by
$\kappa '$
comes in  as a factor $\exp [-\Delta \kappa R]$, where $\Delta \kappa =
\kappa '-\kappa $. The value of $R$ is large, thus, even a small
$\Delta \kappa $ can produce an exponential error in the amplitude.
Using the perturbation-theory regime estimate of $R$ we obtain the error
factor of $\exp [-2(\Delta \kappa /\kappa ) \sqrt{2n}]$ for the
detachment rate.

\subsection{Detachment rates}\label{finres}

The differential $n$-photon detachment rate for the electron in the
initial state $lm$ is obtained from Eqs. (\ref{difprob}) and (\ref{Asad3})
after integration over $\varphi $ and $p$,
\begin{eqnarray}\label{res}
\frac{dw_n}{d\Omega }&=&\frac{pA^2}{4\pi}\left( \frac{\kappa}
{\omega }\right) ^{2\nu }2^\nu \Gamma ^2(1+\nu /2)(2l+1)\frac{(l-|m|)!}
{(l+|m|)!}\left| P_l^{|m|}\left( \sqrt{1+p^2\sin ^2\theta /\kappa ^2}
\right) \right| ^2\nonumber \\
&\times &\left| \sum_{\mu =1,2}(\pm )^{l+m}\frac{(c_\mu +is_\mu )^n}
{\sqrt{2\pi (-iS''_\mu )^{\nu +1}}}\exp \left[ -ic_\mu (\xi +zs_\mu )
\right] \right| ^2 ~,
\end{eqnarray}
where $p=\sqrt{2(n\omega -F^2/4\omega ^2+E_0)}$ is the photoelectron
momentum determined by the energy conservation, and $\nu =0$ for negative
ions. According to the symmetry properties of $A_{{\bf p}n}$, the
differential $n$-photon detachment rate is exactly zero at $\theta =\pi /2$
for odd $n+l+m$.

The total $n$-photon detachment rate of the $lm$ state is obtained
by integrating (\ref{res}),
\begin{equation}\label{rest}
w_n^{(lm)}=2\pi \int _0^\pi \frac{dw_n}{d\Omega } \sin \theta d\theta ~,
\end{equation}
and if we are interested in the total detachment rate for a closed
shell, the sum over $m$ and the electron spin projections must be
completed,
\begin{equation}\label{wallm}
w_n=2\sum _{m=-l}^{l} w_n^{(lm)}~.
\end{equation}
The dominant contribution to this sum is given by the $m=0$ state, since
it is extended along the direction of the field, see Sec. \ref{lowen}.

It is very easy to take the effect of fine-structure splitting into
account. The two fine-structure components $j=l\pm \frac{1}{2}$ of a closed
shell are characterized by different binding energies $|E_0|$ and values of
$\kappa $. The $n$-photon detachment rate for the $j$ sublevel
is then given by
\begin{equation}\label{wj}
\frac{dw_n^{(j)}}{d\Omega }=\frac{2j+1}{2l+1}\sum _{m=-l}^{l}
\frac{dw_n}{d\Omega}~,
\end{equation}
which is exactly what one would expect from naive statistical
considerations.

Of course, one can easily obtain the total detachment rate by summing
the $n$-photon rates over $n$. The smallest $n$ is given by the
integer part of $[(|E_0|+F^2/4\omega ^2)/\omega ]+1$.

\subsection{Limits}

There are two limits which can be usefully explored with the help of
Eq. (\ref{res}). The first is the perturbation theory limit, where the
detachment rate is proportional to the $n$th power of the photon flux
$J=cF^2/(8\pi \omega )$, and the process is described by the generalized
$n$-photon cross section
\begin{equation}\label{defsig}
\frac{d \sigma _n}{d\Omega }=\frac{dw_n}{d\Omega }J^{-n}~.
\end{equation}
The other is the low photoelectron energy limit studied earlier in
\cite{keldysh,perelom}. It enables one to recover the static-field results
\cite{demkov,smch}.

\subsubsection{Perturbation theory limit.}\label{pertth}

To obtain the perturbation-theory limit, it is convenient to re-write
the saddle-point equation (\ref{smu}) for $\sin \omega t_\mu $ in the
following form:
\begin{equation}\label{sinper}
s_\mu=\frac{\omega }{F}\left( p_\parallel \pm i \sqrt{\kappa ^2
+p_\perp ^2}\right)~,
\end{equation}
where $p_\parallel =p\cos \theta $ is the momentum component parallel
to the field. The weak-field regime $\gamma \gg 1$ infers
$|s_\mu |\gg 1$, hence we obtain for $\cos \omega t_\mu $
\begin{equation}\label{cosper}
c_\mu=\pm \sqrt{1+s_\mu ^2}\simeq
-is_\mu +\frac{i}{2s_\mu }+O(s_\mu ^{-2})~.
\end{equation}
Using (\ref{sinper}) and (\ref{cosper}) to calculate the amplitude
(\ref{Asad3}) for $\nu=0$, and retaining only the leading term in $c_\mu $
everywhere, except in $c_\mu +is_\mu $, where the second term is necessary,
we arrive at the following $n$-photon detachment cross section
\begin{eqnarray}\label{sign}
\frac{d \sigma _n}{d\Omega }&=&\frac{pA^2\omega }{4\pi ^2\sqrt{2n\omega }}
(2l+1)\frac{(l-|m|)!}{(l+|m|)!}\left| P_l^{|m|}
\left( \sqrt{1+p_\perp ^2/\kappa ^2}\right) \right| ^2
\left( \frac{\pi e}{nc\omega ^2}\right) ^n\nonumber \\
&\times &\frac{\exp (p_\parallel ^2/\omega )}{\sqrt{\kappa ^2+p_\perp ^2}}
\left[ 1+(-1)^{n+l+m}\cos \Xi \right] ~,
\end{eqnarray}
where $p=\sqrt{2n\omega -\kappa ^2}$, $c\approx 137$ is the speed of
light, $e=2.71\dots $, and $\Xi $ is the momentum-dependent
contribution to the relative phase of the two saddle-point terms
in the amplitude,
\begin{equation}\label{Xi}
\Xi=(2n+1)\tan ^{-1}
\frac{p_\parallel }{\sqrt{\kappa ^2+p_\perp ^2}}+
\frac{p_\parallel \sqrt{\kappa ^2+p_\perp ^2}}{\omega }~.
\end{equation}
This phase varies with the ejection angle of the photoelectron from
$\Xi _0=(2n+1)\tan ^{-1} (p/\kappa )+p\kappa /\omega $ to
$-\Xi _0$, and can be quite large, even for the lowest
$n$ process, $p\sim \sqrt{\omega }$, $\Xi _0\sim \sqrt{n}$,
thus producing oscillations in the photoelectron angular distribution.
Note that in accordance with the general symmetry properties, the cross
section is zero at $\theta =\pi /2$, when $n+l+m$ is odd.

\subsubsection{Low photoelectron energies and the static field
limit.}\label{lowen}

Another simplification of the general formula (\ref{res}) is achieved when
the energy of the photoelectron is low compared to the
binding energy, $p^2\ll \kappa ^2$. Then, following \cite{keldysh,perelom}
one can expand the action $S(\phi _\mu)$ and other quantities calculated
at the saddle points in powers of $p$ up to the 2nd order. For $\nu =0$
which corresponds to the multiphoton detachment from a negative ion, we
obtain
\begin{eqnarray}\label{wperel}
\frac{dw_n}{d\Omega }&=&\frac{pA^2\omega \gamma }{2\pi |E_0|
\sqrt{1+\gamma ^2}}\,\frac{1}{(2^{|m|}|m|!)^2}\,
\frac{2l+1}{4\pi }\,\frac{(l+|m|)!}{(l-|m|)!}
\nonumber \\
&\times &\exp \left\{ -2\frac{|E_0|}{\omega }
\left[ \left( 1+\frac{1}{2\gamma ^2} \right) \sinh ^{-1}\gamma
-\frac{\sqrt{1+\gamma ^2}}{2\gamma }\right] \right\}
\exp \left[-\left( \sinh ^{-1}\gamma -\frac{\gamma }
{\sqrt{1+\gamma ^2}}\right)\frac{p^2}{\omega }\right]\nonumber \\
&\times &\exp \left( -\frac{\gamma p^2\sin ^2\theta }{\omega
\sqrt{1+\gamma ^2}}\right)
\left( \frac{p\sin \theta }{\kappa }\right) ^{2|m|}
\left[ 1+(-1)^{n+l+m}\cos \left(
\frac{2\kappa p \cos \theta \sqrt{1+\gamma ^2} }{\omega \gamma }
\right) \right]~.
\end{eqnarray}
This formula coincides Eq. (53) of Ref. \cite{perelom}.
The $\cos (\dots )$ in the last square brackets of Eq. (\ref{wperel})
appears due to
the interference between the contributions of the two saddle points in
amplitude (\ref{Asad3}), and is the analogue of $\cos \Xi $ in Eq.
(\ref{sign}). It determines the oscillatory behaviour of the angular
dependence of the $n$-photon detachment rate, which would otherwise simply
peak along the direction of the field, $\theta =0$, or $\theta =\pi $,
for $m=0$.

Formula (\ref{wperel}) also shows clearly that the detachment rate for
the states with $m\neq 0$ is much smaller than that of $m=0$, due to the
factor $(p\sin \theta /\kappa )^{2|m|}$. It comes from the leading
term in the expansion of the associated Legendre polynomial
$P_l^{|m|}(x)$ in Eq. (\ref{res}) at $x\approx 1$.

As shown by Perelomov {\em et al.} \cite{perelom}, in the limit
$\omega \rightarrow 0$, Eq. (\ref{wperel}) allows one to recover the
well-known formula for the ionization rate in the static
electric field $F$ \cite{smch},
\begin{equation}\label{static}
w_{\rm stat}=\frac{A^2}{2\kappa ^{2\nu -1}}\,
\frac{(2l+1)(l+|m|)!}{2^{|m|}|m|!(l-|m|)!}
\left( \frac{2F_0}{F}\right) ^{2\nu -|m|-1}
\exp \left( -\frac{2F_0}{3F}\right)
\end{equation}
for negative ion case $\nu =0$. It has been shown recently \cite{fab93}
that the account of the polarization potential $-\alpha e^2/2r^4$
acting between the outer electron and the atomic residue in the negative
ion, changes the numerical pre-exponential factor in Eq. (\ref{static}).
However, this correction is not very large, e.g., it increases
the detachment rate for Ca$^-$ by a factor of 2, in spite of the
large polarizability $\alpha ({\rm Ca})=170$ a.u.

It is worth noting that the perturbation theory formula (\ref{sign})
and the low electron energy limit (\ref{wperel}) have a common range
of applicability. If we use $p\ll \kappa $ in the first, and take the
perturbation theory limit $\gamma \gg 1$ in the second, the two formulae
yield identical results.

\section{Numerical results and discussion}\label{ex}

In this section we use the formulae we obtained within the adiabatic theory
to calculate the photodetachment rates, EPD spectra and photoelectron
angular distributions for H$^-$ and halogen negative ions. These are
so far the most studied species, which enables us to make comparisons with
results of other calculations. Our aim is to show that our theory
achieves good accuracy in describing multiphoton detachment in both
perturbative and strong-field regimes.

To apply the theory, all we need is the asymptotic parameters
$A$ and $\kappa $ of the corresponding bound-state wave functions. The
values of $A$ are tabulated in various sources, and we use those from
\cite{nikitin}. The values of $\kappa $ are calculated using the
corresponding binding energies, $\kappa =\sqrt{2|E_0|}$. They are taken
from the electron affinity tables \cite{hotop}, or obtained by combining
those with the fine-structure intervals of the atomic ground states
\cite{R&S}, when we consider the detachment of $p_{1/2}$ electrons
from the halogens.

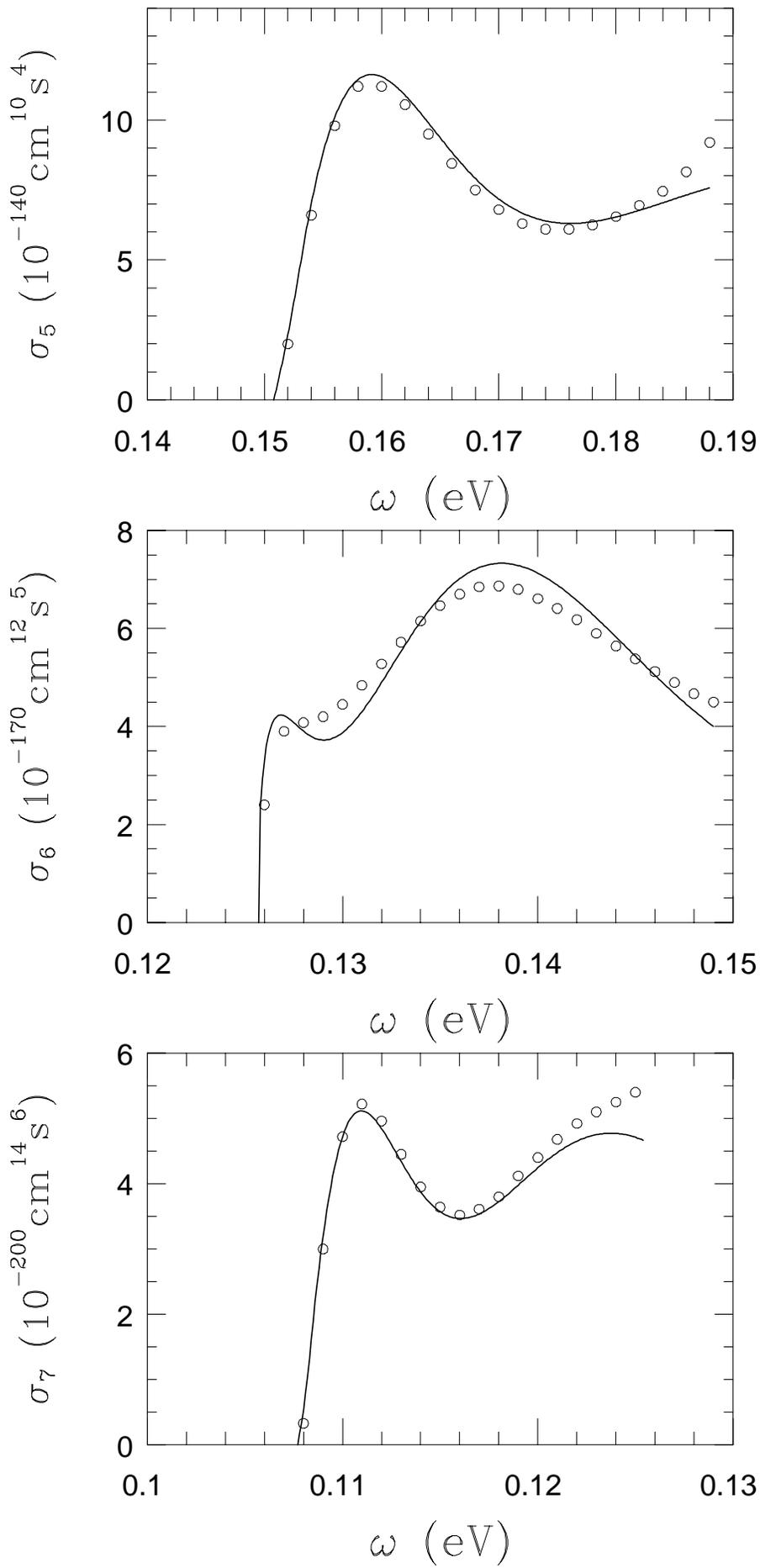
\begin{figure}
\caption{Frequency dependence of the generalized $n$-photon detachment
cross sections for H$^-$, $n=5,~6,~7$. Solid curve: present calculation,
Eq. (\protect\ref{sign}) integrated over angles; open circles:
perturbation theory calculations of Laughlin and Chu
\protect\cite{chu}.\label{sig567}}
\end{figure}

In Fig. \ref{sig567} we present the generalized $n$-photon detachment
cross sections for H$^-$ obtained by integrating the differential
cross sections from Eq. (\ref{sign}) with $A=0.75$ and $\kappa =0.2354$
over $\theta $. The cross section has been multiplied by 2 to account for
the two spin states [cf. Eq. (\ref{wj}) with $l=0$, $j=1/2$].
The results of the perturbation theory calculations
\cite{chu} are shown for comparison. In the latter the interaction of the
electron with the atomic core was described by a model potential which
accounted for the polarizational attraction between the
electron and the atomic core, and was chosen to reproduce the binding
energy of H$^-$, as well as the electron-hydrogen scattering phaseshifts.
Figure \ref{sig567} shows that there is good agreement between
our results and those of \cite{chu}. We checked that even for
$n=3$ the difference does not exceed 20\% at the cross section maximum.

Laughlin and Chu note \cite{chu} that their model-potential results are
close to those obtained in \cite{star} using the hyperspherical method
which accounts for correlations between the two electrons in H$^-$.
They are also in agreement with the two-electron perturbation theory
calculations of \cite{proulx} and the recent $R$-matrix Floquet theory
calculations \cite{dorr}, which also take into account electron
correlations.
The main idea behind those approaches was to reproduce the
negative ion wave function as correctly as possible at all distances,
particularly near the atomic core. This idea is favoured by the experience
gained in a number of problems, such as the single-photon detachment,
electron-atom scattering, etc., where electron correlations are indeed
very important. However, as shown above, the multiphoton problem under
consideration proves to be different. Absorption of several quanta
is dominated by large distances satisfying inequality (\ref{rlarge}).
The complicated behaviour of the wave function inside the atomic core
turns out to be inessential. This is the main reason for the good agreement
we observe in Fig. \ref{sig567}.

To check our theory in the non-perturbative regime, where one must use Eq.
(\ref{res}), the EPD spectra of H$^-$ for the three large field intensities
$I=10^{10}$, $5\times 10^{10}$, and $10^{11}$ W/cm$^2$, of the
10.6-$\mu $m
radiation, $\omega =0.0043$ a.u., are presented in Table \ref{Hmin}. 
For these parameters the electron quiver energy, or the ponderomotive
energy shift, in units of $\omega $, $z=F^2/4\omega ^3=0.894$, 4.472, and
8.945, and the Keldysh parameter $\gamma =1.895$, 0.847, and 0.599,
respectively. For given $\omega $ absorption of a minimum of 7 photons is
required. The ponderomotive
threshold shift changes this number to $n_{\rm min}=8$, 11, and 16.
The calculation of the detachment rates from Eqs. (\ref{res})--(\ref{wj})
has been done using {\em Mathematica} \cite{math}. For the smallest
intensity the lowest EPD peak $n=8$ dominates the total detachment rate,
whereas for the higher intensities many peaks in the EPD spectrum can be
observed. 

The detachment rates in Table \ref{Hmin} are compared with those
obtained in the non-perturbative calculations of Telnov and Chu
\cite{telnov}. They describe their method as a complex-scaling generalized
pseudospectral technique applied to the solution of the time-independent
non-Hermitian Floquet Hamiltonian for the complex quasienergies, and
use the accurate model potential from \cite{chu} to describe the
interaction of the electron with the atomic residue.

There is a good overall agreement between the two calculations. The
discrepancy  usually does not exceed a few per cent, and is slightly
larger for higher EPD peaks and smaller field intensities. The latter is
somewhat puzzling, since there is a good agreement in the
perturbation-theory limit for the 7-photon cross section at
$\omega =0.0043$ a.u.:
\begin{eqnarray}
\sigma _7&=&3.537 \times 10^{-200}~{\rm cm}^{14}{\rm s}^6\quad
\mbox{(Eq. (\ref{sign}) integrated over angles)},\nonumber \\
\sigma _7&=&3.639 \times 10^{-200}~{\rm cm}^{14}{\rm s}^6\quad
\mbox{(result of \cite{telnov})}.\nonumber
\end{eqnarray}

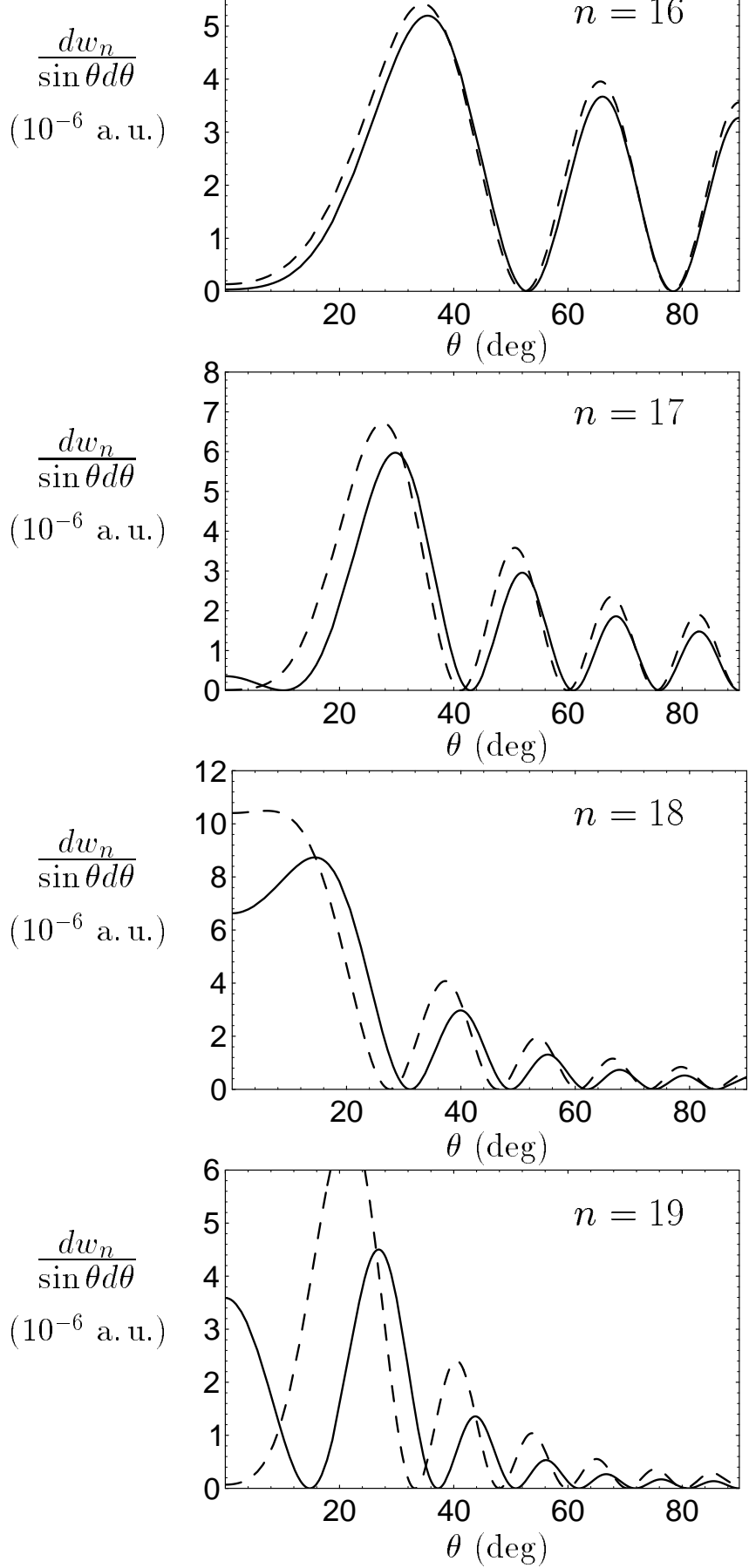
\begin{figure}
\caption{Differential $n$-photon detachment rates of H$^-$ in the strong
laser field, $\omega =0.0043$ a.u., $I=10^{11}$ W/cm$^2$, $z=8.945$,
$\gamma =0.599$. Solid curve: ``exact'' saddle-point calculation,
Eq. (\protect\ref{res}); dashed curve: low photoelectron energy
limit, Eq. (\protect\ref{wperel}). Channels with $n<16$ are closed.
\label{Hdif}}
\end{figure}

In Fig. \ref{Hdif} we show the angular dependence of the photoelectron
peaks for $n=16$, 17, 18, and 19, at  $I=10^{11}$ W/cm$^2$. We have checked
that their shapes, as well as those for other $n$ and intensities, are
practically identical to the angular distributions presented in Figs. 5--7
of \cite{telnov}. Also shown in  Fig. \ref{Hdif} are the differential
detachment rates obtained from Eq. (\ref{wperel}). It works quite well
for two lowest $n$, but the agreement becomes poor with the increase of
the photoelectron energy, e.g., for $n=19$, where $p/\kappa \approx 0.75$.

It is worth stressing again that the remarkable oscillatory
behaviour is caused by the interference of the two saddle-point
contributions in Eq. (\ref{res}), or in other words, the interference
between the electron waves emitted at the two instants separated by
$T/2$, when the field reaches its maximum. The geometrical phase difference
which determines the oscillations of $\cos (\dots )$ in Eq. (\ref{wperel})
can be calculated classically.
Suppose that the electron is considered free at the moment when it
escapes the atomic particle. Its classical coordinate is then given by
${\bf r}(t)=\int ^t{\bf k}_{t'}dt'=({\bf F}/\omega ^2)\cos \omega t$. At
the two instants $t_\mu $ when the adiabatic transition takes place
we have
\begin{displaymath}
{\bf r}(t_\mu )=\pm \frac{{\bf F}}{\omega ^2}\sqrt{1+\gamma ^2}
=\pm \frac{{\bf F}}{F}\,\frac{\kappa \sqrt{1+\gamma ^2}}{\gamma \omega }~,
\end{displaymath}
where Eq. (\ref{gamma}) is used for small momenta $p\ll \kappa $.
Note that though $t_\mu $ are complex, the corresponding electron
coordinates are real. These points located at the opposite sides of
the atomic particle, are sources of the two electron waves emitted
at the angle $\theta $ with respect to ${\bf F}$. The geometrical phase
is obtained by multiplying the base $|{\bf r}(t_1)-{\bf r}(t_2)|$ by the
projection of the electron momentum on the the direction of the field
$p\cos \theta $.

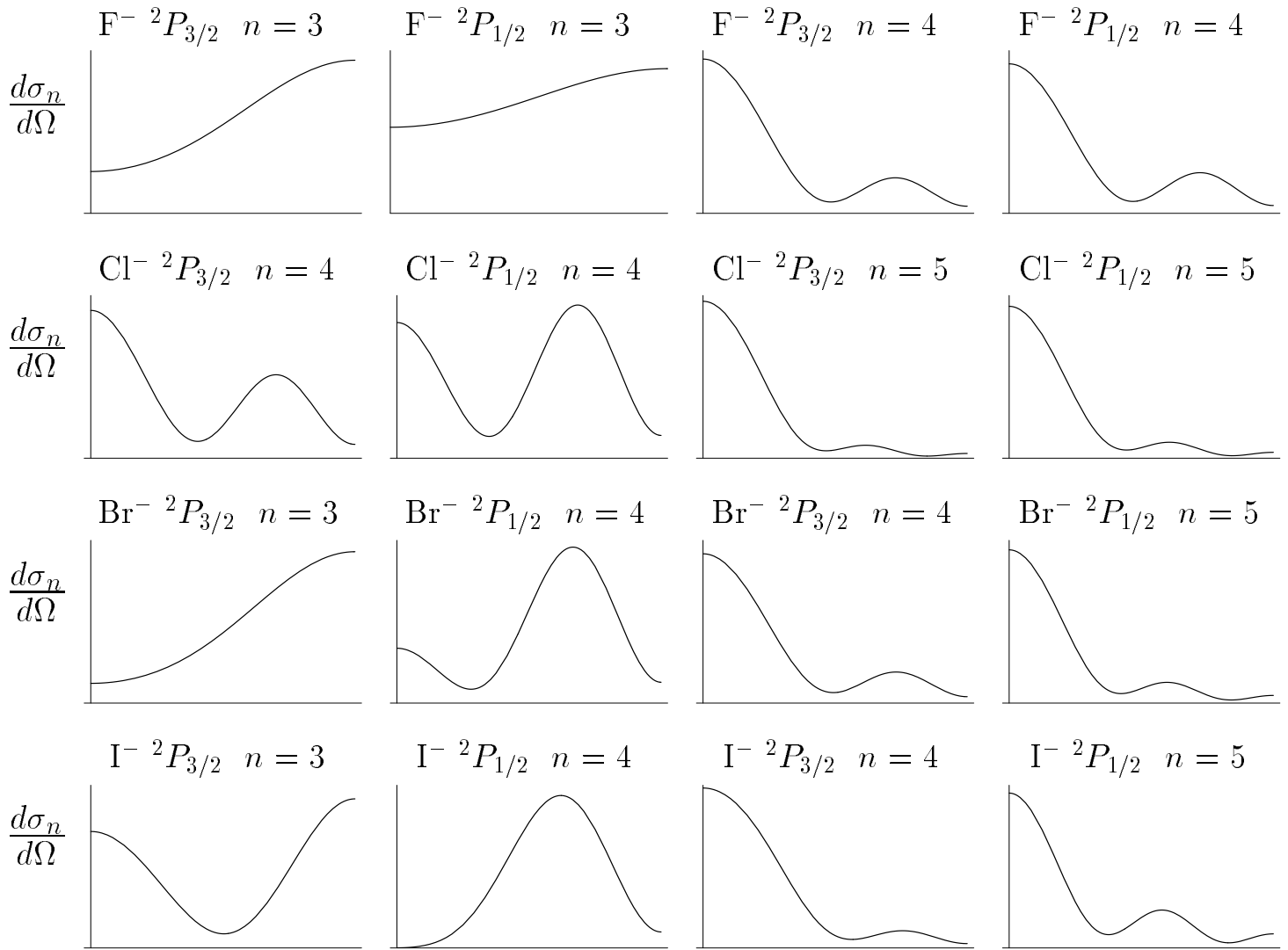
\begin{figure}
\caption{Differential $n$-photon cross sections for the electron detachment
from the halogen negative ions, which leaves the atom in the $^2P_{3/2}$ or
$^2P_{1/2}$ states, Eqs. (\protect \ref{sign}), (\protect \ref{wj}).
\label{halfig}}
\end{figure}

Our results for halogen negative ions are presented in Fig.
\ref{halfig} and Table \ref{hal}. They have been obtained from Eqs.
(\ref{sign}), (\ref{wj}) for comparison with the perturbation theory
calculations \cite{crance} at the Nd:YAG laser frequency
$\omega =0.0428$ a.u. In that work the non-relativistic Hartree-Fock wave
functions of the valence $np$ electrons were used, together with
experimental threshold energies. The photoelectron was described in the
plane wave approximation. This approximation is equivalent to our use of
the Volkov wave function in the perturbation-theory limit.
As shown in the earlier works by Crance
\cite{crance87}, the multiphoton detachment results obtained in the plane
wave approximation are close to those obtained using the frozen core
Hartree-Fock wave functions of the photoelectron.

The shapes of angular distributions presented in Fig. \ref{halfig} are
quite close to those in Fig. 2 of Ref. \cite{crance}, although
quantitative comparison is not feasible due to the use of an arbitrary
vertical scale in \cite{crance}.

The absolute values of the $n$-photon detachment cross sections in
from our calculations and \cite{crance} compare reasonably on a
logarithmic scale for all cases shown in Table \ref{hal}. However,there is
a systematic discrepancy. To find its origin let us recall that the
multiphoton detachment rate is very sensitive to the asymptotic behaviour
of the bound-state wave function (see end of Sec. \ref{adap}). In
\cite{crance} the Hartree-Fock wave functions have been used. Their
asymptotic behaviour $\exp (-\kappa _{\rm hf}r)$ is different from the
correct $\exp (-\kappa r)$, based on the experimental value of $\kappa $.
Thus, to account for the discrepancy in Table \ref{hal}, the Hartree-Fock
based results should be multiplied by the factor
\begin{equation}\label{corr}
\sim \exp [2(\kappa _{\rm hf}-\kappa )R]~,
\end{equation}
where, according to (\ref{rlarge}), $R\approx 1/\sqrt{\omega }$.
Formula (\ref{corr}) shows that when $\kappa _{\rm hf}>\kappa $, the
Hartree-Fock based calculations underestimate the detachment rate, while
for $\kappa _{\rm hf}<\kappa $ they overestimate it.

The Hartree-Fock values of $\kappa _{\rm hf}$ are 0.602, 0.545,
0.528, and 0.508, for the outer $np$ subshell of F$^-$, Cl$^-$, Br$^-$,
and I$^-$, respectively. Examination of the lowest $n$ cross sections
throughout Table \ref{hal} shows that the qualitative explanation of
the discrepancy based on (\ref{corr}) is correct. For example,
for F$^-$ where $\kappa _{\rm hf}=0.6$ and
$\kappa =0.5$, formula  (\ref{corr}) gives 2.6, whereas the ratio of the
3-photon detachment cross sections for F$^-$, $j=3/2$, in Table \ref{hal}
is 4.3. Also, the best agreement in Table \ref{hal} is achieved for
Br$^-$, $j=1/2$, where $\kappa _{\rm hf}$ is very close to the correct
value. Therefore, we conclude that the incorrect asymptotic behaviour of
the Hartree-Fock wave functions can produce significant errors in the
multiphoton detachment rates. This must be kept in mind when comparisons
are made between different $n$-photon detachment calculations
\cite{hart96}.

\section{Summary}

The main result of our work is that the adiabatic theory approach to the
multiphoton problems originally suggested by Keldysh, is more powerful
and accurate than is generally believed. It yields accurate multiphoton
detachment rates for negative ions, and reveals a number of interesting
details about the physics of the problem: the role of large distances
and asymptotic behaviour of the bound-state wave function, and the origin
of oscillations in the angular distribution of photoelectrons.
The formulae obtained in the paper allow one to make simple and reliable
estimates of the $n$-photon detachment rates for $n>2$ in both
perturbative and non-perturbative regimes.

\acknowledgments

We would like to thank V. V. Flambaum, W. R. Johnson and O. P. Sushkov
for useful discussions, and V. N. Ostrovsky for providing us with a
reference. This work has been supported by the Australian Research Council.

\appendix

\section{Calculation of transition rates in a strong periodic field}
\label{prob}
Suppose the system is in the initial state
\begin{displaymath}
\psi _0(t)=e^{-iE_0t}\phi _0~, \qquad H_0\phi _0=E_0\phi _0~
\end{displaymath}
of the time-independent Hamiltonian $H_0$, and a periodic field
$V(t)=V(t+T)$ is turned on adiabatically. We assume that this field can
be strong, so that the lowest-order perturbation theory is inapplicable.
The time-dependent wave function of the system
\begin{equation}\label{timedS}
i\frac{\partial \Psi }{\partial t}=[H_0+V(t)]\Psi 
\end{equation}
can be presented as the sum
\begin{equation}\label{sumlam}
\Psi (t)=\psi _0(t)+\sum _\lambda a_\lambda (t)\psi _\lambda (t)
\end{equation}
over the set of eigenstates $\psi _\lambda (t)$ of the total Hamiltonian,
\begin{displaymath}
i\frac{\partial \psi _\lambda }{\partial t}=[H_0+V(t)]\psi _\lambda 
\end{displaymath}
which represent the possible final states of the system, $a_\lambda (t)$
being the amplitude of finding the system in one of these states. 
In Eq. (\ref{sumlam}) we assume that $a_\lambda (t)\rightarrow 0$
at $t\rightarrow -\infty $, and the rate of the transition
$\psi _0 \rightarrow \psi _\lambda $ is given by
$d|a_\lambda (t)|^2/dt$.

According to the Floquet theorem each state
$\psi _\lambda (t)=e^{-iE_\lambda t}\phi _\lambda (t)$
is characterized by its quasienergy $E_\lambda $ and the corresponding
periodic quasienergy wave function $\phi _\lambda (t)=\phi _\lambda (t+T)$,
found from
\begin{displaymath}
i\frac{\partial \phi _\lambda }{\partial t}=[H_0+V(t)-E_\lambda ]
\phi _\lambda ~.
\end{displaymath}
At any given $t$ the quasienergy wave functions form a complete orthonormal
set, $\langle \psi _\lambda |\psi _{\lambda '}\rangle =
\langle \phi _\lambda |\phi _{\lambda '}\rangle =
\delta _{\lambda \lambda '}$.

After inserting $\Psi (t)$ (\ref{sumlam}) into Eq. (\ref{timedS})
and projecting it onto the state $\langle \psi _\lambda (t)|$, we arrive at
\begin{eqnarray}\label{dadt}
\frac{da_\lambda }{dt}=-i\langle \psi _\lambda (t)|V(t)|\psi _0(t)
\rangle =
-ie^{iE_\lambda t}e^{-iE_0 t}\langle \phi _\lambda (t)|V(t)|
\phi _0\rangle ~.
\end{eqnarray}
The last matrix element is a periodic function of time,
\begin{equation}\label{perfun}
\langle \phi _\lambda (t)|V(t)|\phi _0\rangle =\sum _ne^{-i\omega nt}
A_{\lambda n}~,
\end{equation}
where $\omega =2\pi /T$, and
\begin{equation}\label{AFour}
A_{\lambda n}=\frac{1}{T}\int _0^T\langle \phi _\lambda (t)|V(t)|
\phi _0\rangle e^{i\omega nt}dt~.
\end{equation}
Using (\ref{perfun}) we re-write (\ref{dadt}) as
\begin{displaymath}
\frac{da_\lambda }{dt}=-i\sum _ne^{i(E_\lambda -E_0-nw)}A_{\lambda n} ~,
\end{displaymath}
and find
\begin{displaymath}
a_\lambda =\int ^t\frac{da_\lambda }{dt}=
-\sum _n \frac{e^{i(E_\lambda -E_0-nw)}e^{\eta t}}{E_\lambda -E_0-nw-i
\eta }A_{\lambda n}
\end{displaymath}
where the energies $E_\lambda $ have been given an infinitesimal
shift $-i\eta $ to make $\int ^t\dots dt$ converge at $-\infty $.
The probability is given by
\begin{displaymath}
|a_\lambda |^2=\sum _n \frac{e^{2\eta t}|A_{\lambda n}|^2}
{(E_\lambda -E_0-nw)^2+\eta ^2}+{\mbox{oscillating terms}}~,
\end{displaymath}
and the rate is
\begin{displaymath}
\frac{d}{dt}|a_\lambda (t)|^2=\sum _n \frac{2\eta e^{2\eta t}}
{(E_\lambda -E_0-nw)^2+\eta ^2}|A_{\lambda n}|^2 ~,
\end{displaymath}
where we dropped the oscillating terms since they do not contribute to
the transition rate after we average it over a period.
Finally, we take the limit $\eta \rightarrow 0$ using the following
representation of the $\delta $-function,
\begin{displaymath}
\lim _{\eta \rightarrow 0}\frac{2\eta }{x^2+\eta ^2}=2\pi \delta (x)
\end{displaymath}
and obtain
\begin{equation}\label{finform}
\frac{d}{dt}|a_\lambda (t)|^2=2\pi\sum _n |A_{\lambda n}|^2 \delta
(E_\lambda -E_0-nw)~,
\end{equation}
where the amplitude $A_{\lambda n}$ given by Eq. (\ref{AFour}) can be
written as
\begin{equation}\label{Afin}
A_{\lambda n}=\frac{1}{T}\int _0^T\langle \psi _\lambda (t)|V(t)|
\psi _0(t)\rangle dt~,
\end{equation}
due to the energy conservation $E_\lambda -E_0=nw$ implied by the
$\delta $-function. This amplitude describes the $n$-quantum process,
and the total transition rate (\ref{finform}) is the sum over all
such processes. If the spectrum of $\lambda $ is continuous, the
differential transition rate $dw_\lambda $ is proportional to the
corresponding density of states,
\begin{equation}\label{fincont}
dw_\lambda =2\pi\sum _n |A_{\lambda n}|^2 \delta (E_\lambda -E_0-nw)
d\rho _\lambda ~.
\end{equation}

\section{Saddle-point method for integrals with a singularity}
\label{apsad}

Consider the integral
\begin{equation}\label{J}
J=\int \limits_{C}g(x)\exp [-\lambda f(x)]dx
\end{equation}
for $\lambda \rightarrow \infty $. In this case it is well known
\cite{method} that the integration contour $C$ should be deformed to go
through the saddle point $x_0$ where $f'(x_0)=0$. The vicinity of this
point gives the main contribution to the integral. If the function
$g(x)$ is not singular at $x=x_0$, the integral (\ref{J}) is evaluated
as
\begin{equation}\label{int2}
J\simeq g(x_0)\sqrt{\frac{2\pi}{\lambda f''(x_0)}}\exp [-\lambda f(x_0)]~.
\end{equation}

If $g(x)$ has a singularity at $x=x_0$, e.g. $g(x)=(x-x_0)^{-\nu }$, the
saddle-point answer has to be modified. Consider the following integral,
\begin{equation}\label{int3}
J_\nu =\int \frac{\exp [-\lambda f(x)]}{(x-x_0)^\nu}dx~.
\end{equation}
By using the transformation \cite{ryzhik}
\begin{equation}\label{repr}
\frac{1}{(x-x_0)^\nu}=\frac{1}{\Gamma (\nu )}
\int _0^\infty d\xi \xi ^{\nu -1}\exp [-\xi (x-x_0)]~,
\end{equation}
we turn (\ref{int3}) into the double integral,
\begin{equation}\label{int4}
\int _0^\infty d\xi \xi ^{\nu -1}\int \exp [-\lambda f(x)-\xi (x-x_0)]dx~.
\end{equation}
Calculating $\int \dots dx$ by means of (\ref{int2}) and
then integrating over $\xi $ we obtain for $\lambda \rightarrow \infty  $,
\begin{equation}
J_\nu \simeq i^\nu \,
\frac{\Gamma (\nu /2)}{2\Gamma (\nu)}\,\sqrt{\frac{2\pi}
{\lambda f''(x_0)}}\,
[2\lambda f''(x_0)]^{\nu /2}\exp [-\lambda f(x_0)]~.
\end{equation}
For $\nu \rightarrow 0$ we, of course, recover (\ref{int2}).


\tightenlines

\newcommand{\ti}{\times }
\begin{table}
\caption{The EPD spectra of H$^-$ in the strong laser field of
$\omega =0.0043$ a.u. The detachment rates calculated by our saddle-point
method (SP), Eq. (\protect \ref{res}), $A=0.75$ and $\kappa =0.235$, are
compared with the non-perturbative results by Telnov and Chu
\protect \cite{telnov}.\label{Hmin}}

\begin{tabular}{ccccccc}
& \multicolumn{6}{c}{$n$-photon detachment rate (a.u.)} \\
\cline{2-7}
& \multicolumn{2}{c}{$I=10^{10}$ W/cm$^2$}&
 \multicolumn{2}{c}{$I=5\ti 10^{10}$ W/cm$^2$} &
 \multicolumn{2}{c}{$I=10^{11}$ W/cm$^2$} \\
$n$ & SP & \cite{telnov} & SP & \cite{telnov} & SP &
\cite{telnov} \\
\tableline
8 & $6.69\ti 10^{-10}$ &$7.12\ti 10^{-10}$ &$-$ &$-$ &$-$ &$-$ \\
9 & $1.92\ti 10^{-10}$ &$2.03\ti 10^{-10}$ &$-$ &$-$ &$-$ &$-$ \\
10& $4.08\ti 10^{-11}$ &$4.32\ti 10^{-11}$ &$-$ &$-$ &$-$ &$-$ \\
11& $4.99\ti 10^{-12}$ &$5.26\ti 10^{-12}$ &$5.44\ti 10^{-7}$ &
$4.07\ti 10^{-7}$ &$-$ &$-$ \\
12& $7.24\ti 10^{-13}$ &$7.86\ti 10^{-13}$ &$4.68\ti 10^{-7}$ &
$4.88\ti 10^{-7}$ &$-$ &$-$ \\
13& $2.03\ti 10^{-13}$ &$2.27\ti 10^{-13}$ &$3.57\ti 10^{-7}$ &
$3.69\ti 10^{-7}$ &$-$ &$-$ \\
14&&&$1.24\ti 10^{-7}$ &$1.30\ti 10^{-7}$ &$-$ &$-$ \\
15&&&$9.54\ti 10^{-8}$ &$9.72\ti 10^{-8}$ &$-$ &$-$ \\
16&&&$8.28\ti 10^{-8}$ &$8.52\ti 10^{-8}$ &$4.31\ti 10^{-6}$ &
$4.32\ti 10^{-6}$\\
17&&&$4.72\ti 10^{-8}$ &$4.88\ti 10^{-8}$ &$3.09\ti 10^{-6}$ &
$3.14\ti 10^{-6}$\\
18&&&$1.99\ti 10^{-8}$ &$2.06\ti 10^{-8}$ &$2.55\ti 10^{-6}$ &
$2.48\ti 10^{-6}$\\
19&&&$7.59\ti 10^{-9}$ &$7.87\ti 10^{-9}$ &$1.24\ti 10^{-6}$ &
$1.24\ti 10^{-6}$\\
20&&&$3.73\ti 10^{-9}$ &$3.94\ti 10^{-9}$ &$1.28\ti 10^{-6}$ &
$1.22\ti 10^{-6}$\\
21&&&$2.71\ti 10^{-9}$ &$2.92\ti 10^{-9}$ &$1.01\ti 10^{-6}$ &
$1.01\ti 10^{-6}$\\
22&&&$2.18\ti 10^{-9}$ &$2.37\ti 10^{-9}$ &$4.99\ti 10^{-7}$ &
$5.05\ti 10^{-7}$\\
23&&&$1.62\ti 10^{-9}$ &$1.77\ti 10^{-9}$ &$3.74\ti 10^{-7}$ &
$3.64\ti 10^{-7}$\\
24&&&$1.09\ti 10^{-9}$ &$1.18\ti 10^{-9}$ &$4.37\ti 10^{-7}$ &
$4.25\ti 10^{-7}$\\
25&&&$6.62\ti 10^{-10}$ &$7.17\ti 10^{-10}$ &$4.32\ti 10^{-7}$ &
$4.28\ti 10^{-7}$\\
26&&&$3.72\ti 10^{-10}$ &$4.02\ti 10^{-10}$ &$3.34\ti 10^{-7}$ &
$3.34\ti 10^{-7}$\\
27&&&$1.95\ti 10^{-10}$ &$2.10\ti 10^{-10}$ &$2.11\ti 10^{-7}$ &
$2.12\ti 10^{-7}$\\
28&&&$9.69\ti 10^{-11}$ &$1.04\ti 10^{-10}$ &$1.16\ti 10^{-7}$ &
$1.17\ti 10^{-7}$\\
29&&&&&$6.23\ti 10^{-8}$ &$6.26\ti 10^{-8}$\\
30&&&&&$3.88\ti 10^{-8}$ &$3.94\ti 10^{-8}$\\
31&&&&&$3.17\ti 10^{-8}$ &$3.26\ti 10^{-8}$\\
32&&&&&$3.02\ti 10^{-8}$ &$3.14\ti 10^{-8}$\\
33&&&&&$2.90\ti 10^{-8}$ &$3.03\ti 10^{-8}$\\
34&&&&&$2.62\ti 10^{-8}$ &$2.75\ti 10^{-8}$\\
\tableline
Sum &$9.07\ti 10^{-10}$ & $9.66\ti 10^{-10}$ &
$1.76\ti 10^{-6}$ &$1.67\ti 10^{-6}$ &
$1.61\ti 10^{-5}$ &$1.61\ti 10^{-5}$ \\
\end{tabular}
\end{table}


\begin{table}
\caption{Comparison of the $n$-photon detachment cross sections from the
halogen negative ions obtained by the saddle-point method (SP), Eqs.
(\protect \ref{sign}), (\protect \ref{wj}), with the perturbation theory
calculations by Crance \protect \cite{crance} at $\omega =0.0428$ a.u.
For each $n$, $\log \sigma _n^{(j)}$ is shown, $\sigma _n^{(j)}$ being in
units of cm$^{2n}$s$^{n-1}$; $j=3/2$ and $1/2$ for the
$^2P_{3/2}$ and $^2P_{1/2}$ final states of the atom.\label{hal}}
\begin{tabular}{ccccccc}
Ion and its & & \multicolumn{2}{c}{$\log \sigma _n^{(3/2)}$} & &
\multicolumn{2}{c}{$\log \sigma _n^{(1/2)}$} \\
\cline{3-4} \cline{6-7}
parameters & $n$ & SP & \cite{crance} & $n$ & SP & \cite{crance}\\
\tableline
Fluorine             & 3 & $-81.62$ & $-82.25$ & 3 & $-82.01$ & $-83.21$ \\
 $A=0.7$             & 4 & $-113.45$ & $-114.06$ & 4 & $-113.81$ &
$-114.39$ \\
$\kappa _{3/2}=0.4998$& 5 & $-145.36$ & $-145.87$ & 5 & $-145.71$ &
$-146.21$ \\
$\kappa _{1/2}=0.5035$& 6 & $-177.40$ & $-177.75$ & 6 & $-177.74$ &
$-178.08$ \\
\tableline
Chlorine             & 4 & $-113.14$ & $-113.42$ & 4 & $-113.53$ &
$-113.74$ \\
 $A=1.3$             & 5 & $-145.05$ & $-145.26$ & 5 & $-145.47$ &
$-145.64$ \\
$\kappa _{3/2}=0.5156$& 6 & $-177.05$ & $-177.12$ & 6 & $-177.45$ &
$-177.49$ \\
$\kappa _{1/2}=0.5233$& 7 & $-209.14$ & $-209.08$ & 7 & $-209.53$ &
$-209.44$ \\
\tableline
Bromine              & 3 & $-80.99$ & $-81.23$ & 4 & $-113.52$ & $-113.52$
\\
 $A=1.4$             & 4 & $-112.81$ & $-113.06$ & 5 & $-145.51$ &
$-145.46$ \\
$\kappa _{3/2}=0.4973$& 5 & $-144.73$ & $-144.85$ & 6 & $-177.48$ &
$-177.31$ \\
$\kappa _{1/2}=0.5300$& 6 & $-176.77$ & $-176.75$ & 7 & $-209.55$ &
$-209.25$ \\
\tableline
Iodine               & 3 & $-80.59$ & $-80.85$ & 4 & $-113.35$ &
$-113.10$ \\
 $A=1.8$             & 4 & $-112.27$ & $-112.46$ & 5 & $-145.48$ &
$-145.13$ \\
$\kappa _{3/2}=0.4742$& 5 & $-144.25$ & $-144.31$ & 6 & $-177.45$ &
$-176.98$ \\
$\kappa _{1/2}=0.5423$& 6 & $-176.33$ & $-176.29$ & 7 & $-209.50$ &
$-208.93$ \\
\end{tabular}
\end{table}


\begin{references}
\bibitem{note} Note that the final answer for the probability presented in
\cite{keldysh} [Eq. (16)] is two times smaller than one would get from
(\ref{Asad3}), neglecting the interference between the two terms. This
possibly means that the contribution of only one saddle point has been
taken into account in \cite{keldysh}.
\end{references}
\end{document}